\pdfoutput=1
\documentclass[aip,graphicx,author-year]{revtex4-1}
\usepackage{graphicx}
\usepackage{braket}
\usepackage{amsmath}
\usepackage{hyperref}
\usepackage{float}

\providecommand{\keywords}[1]
{
  \small	
  \textbf{Keywords:} #1
}

\draft 

\begin{document}


\title{Occupation Number Representation of Graph} 



\author{Haoqian Pan}
\email[]{52215500040@stu.ecnu.edu.cn}
\author{Changhong LU}
\author{Ben Yang}
\affiliation{School of Mathematical Sciences, East China Normal University}


\date{\today}

\begin{abstract}
In this paper, we propose a new way to represent graphs in quantum space. In that approach, we replace the rows of the adjacency matrix of the graph by state vectors in the occupation number representation. 
Unlike the traditional definition of graph states, we actually let the occupation number of a single-particle state denote the number of edges between each two adjacent vertices. 
This allows us to avoid taking into account the interaction between each two particles. Based on the creation and annihilation operators, we propose the edge creation and annihilation operators. 
With these two operators, we can implement the fundamental operation of adding and removing edges and vertices in a graph. 
Then all additional operations in the graph such as vertex contractions can be defined. Our method can be used to represent both simple and multigraphs. 
Directed and undirected graphs are also compatible with our approach. 
The method of representation proposed in this paper enriches the theory of graph representation in quantum space. 
\end{abstract}



\maketitle 

\section{Introduction}

\begin{figure}[H]
  \includegraphics[width=3cm]{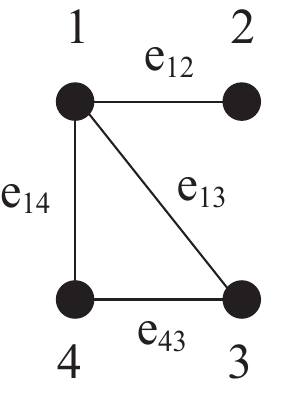}
  \centering
  \caption {A graph with 4 vertices and 4 edges}
  \label{fig:graph}
\end{figure}

In recent years, with the development of quantum computer, more and more NP-Complete (NPC) problems are tried with quantum algorithms. For example, 3-SAT \citep{RN385}, Traveling Salesman Problem (TSP) \citep{RN390}, graph coloring problem \citep{RN389}, independent set problem \citep{RN386}, clique problem \citep{RN387}, vertex cover problem \citep{RN388} and so on. Most of the 
NPC problems are essentially related to graph.  Therefore generating a graph state is crucially in quantum computing, see \cite{RN394}, \cite{RN393}, \cite{RN395}, \cite{RN396} and \cite{RN397}. 
In general, a graph state can be represented by a multi-qubit state, where each vertex is represented by a qubit and for each two adjacent vertices in the graph there is an interaction. 
In this paper, we find that there is another way to represent graphs in the structure of occupation number representation. 
To clarify our idea, let us start with a general mathematical representation of graph. Graph is a structure with vertices and edges identified by $V$ and $E$. As an example, in Fig.~\ref{fig:graph}, there is a graph with 4 vertices and 4 edges. 
Thus, we have $V= \{1,2,3,4\}$ and $E = \{e_{12},e_{13}, e_{14},e_{43}\}$.  
One of the most popular representations of graph is adjacency matrix. The adjacency matrix of $G$ is the $|V| \times |V|$ matrix $M := (m_{i,j})$, where $m_{i,j}$ 
is the number of edges for two adjacent vertices $i$ and $j$ \citep{RN392}. In Fig.~\ref{fig:graph}, the adjacency matrix $M$ is
$$
\begin{bmatrix}
  0 & 1 & 1 & 1 \\
  1 & 0 & 0 & 0 \\
  1 & 0 & 0 & 1 \\
  1 & 0 & 1 & 0 \\
  \end{bmatrix}
$$
By specifying the occupation numbers of the single-particle states in Fock Space, we have the state vector $\ket{n_{1},n_{2},\dots,n_{i},\dots,n_{\infty}}$. 
We find that the state vector can be used to represent the row vector in the adjacency matrix of the graph. 
Such that $[m_{i,1},m_{i,2},\dots,m_{i,|V|}]$ can be represented by $\ket{n_{i,1},n_{i,2},\dots,n_{i,|V|}}$. 
The essence of that representation is that the single-particle state $\ket{\alpha_{i,j}}$ is used to represent the adjacency of vertex $i$ and $j$. Due to the single-particle state, the interaction operator $U^{\{i,j\}}$ used in the classical definition of graph state \citep{RN393} is no longer needed.  
More than that, the number of particles in state $\ket{\alpha_{i,j}}$, $n_{i,j}$, can be seen as the number of edges between $i$ and $j$. 
As we understand, distinguished by the limits of edges between each pair of vertices in graph, there are simple graph and multigraph. 
For simple graph, the maximum number of edges between each two adjacent nodes is one, while for multigraph there is no limit. 
Evidently, systems of Fermions are matched with simple graph, similarly with Bosons for multigraph. We believe the wonderful coincidence can build the bridge between graph and quantum computer. 
Go on with the state vector $\ket{n_{1},n_{2},\dots,n_{i},\dots,n_{\infty}}$, if we apply the creation and annihilation operators to it, 
we have Eq. \ref{creation} and \ref{annihilation} \citep{RN391}
\begin{equation}
  a_{i}^{+} \ket{n_{1},n_{2},\dots,n_{i},\dots,n_{\infty}} = (n_{i} + 1)^{1/2}\ket{n_{1},n_{2},\dots,n_{i} + 1,\dots,n_{\infty}} \label{creation}
\end{equation}
\begin{equation}
  a_{i} \ket{n_{1},n_{2},\dots,n_{i},\dots,n_{\infty}} = (n_{i})^{1/2}\ket{n_{1},n_{2},\dots,n_{i} - 1,\dots,n_{\infty}} \label{annihilation}
\end{equation}
Obviously, creation and annihilation operators can be used to create a graph $G(V,E)$ from vacuum state $\ket{0}$ such as Eq. \ref{generate graph}.
\begin{equation}
\begin{aligned}
  \ket{n_{i,1},n_{i,2},\dots,n_{i,|V|}} & = (n_{i,1}!n_{i,2}!\dots n_{i,|V|}!)^{1/2}(a_{i,1}^{+})^{n_{i,1}}(a_{i,2}^{+})^{n_{i,2}}\\
                                        & \dots(a_{i,|V|}^{+})^{n_{i,|V|}}\ket{0}, \forall i \in V \label{generate graph}
\end{aligned}
\end{equation}
Beyond this, we also find that creation and annihilation operators can be used to implement operations in graphs, as will be discussed in the forthcoming section. 

\section{Operations of graph in occupation number representation}
\subsection{Delete and add edge}\label{section:Delete and add edge}

For example, let $e_{i,j}$ be the edge between node $i$ and $j$ in a simple graph $G(V,E)$. If we want to delete $e_{i,j}$, then we need to set $m_{i,j}$ and $m_{j,i}$ to 0. Due to the properties of simple graph, $m_{i,j}$ and $m_{j,i}$ in adjacency matrix are 1. 
So as $n_{i,j}$ and $n_{j,i}$ in state vector. For simple graph, we only need to apply the annihilation operator to state vectors of row $i$ and row $j$. 
However, in order to be compatible with multigraph, let us consider a more general scenario. Let $k$ be the number of edges between $i$ and $j$ and $d$ be the number of edges waiting for removing. 
Clearly, $d \leq k$ and $k = n_{i,j} = n_{j,i}$. Thus, we have

\begin{equation}
  \begin{aligned}
    (a_{i,j})^{d} \ket{n_{i,1},n_{i,2},\dots,n_{i,j},\dots,n_{i,|V|}} = (k!/(k-d)!)^{1/2}
    \cdot \ket{n_{i,1},n_{i,2},\dots,n_{i,j} - d,\dots,n_{i,|V|}} \label{dij}
  \end{aligned}
  \end{equation}
  \begin{equation}
    \begin{aligned}
      (a_{j,i})^{d} \ket{n_{j,1},n_{j,2},\dots,n_{j,i},\dots,n_{j,|V|}} = (k!/(k-d)!)^{1/2}
      \cdot \ket{n_{j,1},n_{j,2},\dots,n_{j,i} - d,\dots,n_{j,|V|}}\label{dji}
    \end{aligned}
    \end{equation}
By moving the factors of the right side of Eq. \ref{dij} and \ref{dji} to left, we have a new operator $DE_{d}^{i,j}$. See Eq. \ref{edge annihilation operator}
\begin{equation}
  DE_{d}^{i,j} = \frac{(a_{i,j})^{d}}{(k!/(k-d)!)^{1/2}} \label{edge annihilation operator}
\end{equation}
We call $DE_{d}^{i,j}$ is the edge annihilation operator. It can delete $d$ edges between $i$ and $j$. With edge annihilation operator, Eq. \ref{dij} and \ref{dji} can be rewriten as 

\begin{equation}
  \begin{aligned}
    DE_{d}^{i,j}\ket{n_{i,1},n_{i,2},\dots,n_{i,j},\dots,n_{i,|V|}} = \ket{n_{i,1},n_{i,2},\dots,n_{i,j} - d,\dots,n_{i,|V|}} \label{newdij}
  \end{aligned}
\end{equation}
\begin{equation}
    \begin{aligned}
      DE_{d}^{j,i}\ket{n_{j,1},n_{j,2},\dots,n_{j,i},\dots,n_{j,|V|}} = \ket{n_{j,1},n_{j,2},\dots,n_{j,i} - d,\dots,n_{j,|V|}} \label{newdji}
    \end{aligned}
\end{equation}

Talk about the operation of adding edges between $i$ and $j$. 
The process of adding is analogous to that of deleting. In contrast to decreasing the occupation number of $\ket{\alpha_{i,j}}$, adding edges is equivalent to improving the occupation number $n_{i,j}$. 
Follow Eq. \ref{dij} and \ref{dji}, we have 

\begin{equation}
  \begin{aligned}
    (a_{i,j}^{+})^{d} \ket{n_{i,1},n_{i,2},\dots,n_{i,j},\dots,n_{i,|V|}} = ((k + d)!/k!)^{1/2}
    \cdot \ket{n_{i,1},n_{i,2},\dots,n_{i,j} + d,\dots,n_{i,|V|}} \label{aij}
  \end{aligned}
\end{equation}
\begin{equation}
    \begin{aligned}
      (a_{j,i}^{+})^{d} \ket{n_{j,1},n_{j,2},\dots,n_{j,i},\dots,n_{j,|V|}} = ((k + d)!/k!)^{1/2}
      \cdot \ket{n_{j,1},n_{j,2},\dots,n_{j,i} + d,\dots,n_{j,|V|}} \label{aji}
    \end{aligned}
\end{equation}

Also, we can define the edge creation operator $AE_{d}^{i,j}$ 
\begin{equation}
  AE_{d}^{i,j} = \frac{(a_{i,j}^{+})^{d}}{((k + d)!/k!)^{1/2}}  \label{edge creation operator}
\end{equation}

Finally, we get Eq. \ref{newaij} and \ref{newaji}.
\begin{equation}
  \begin{aligned}
    AE_{d}^{i,j}\ket{n_{i,1},\dots,n_{i,j},\dots,n_{i,|V|}} = \ket{n_{i,1},\dots,n_{i,j} + d,\dots,n_{i,|V|}} \label{newaij}
  \end{aligned}
\end{equation}
\begin{equation}
    \begin{aligned}
      AE_{d}^{j,i} \ket{n_{j,1},\dots,n_{j,i},\dots,n_{j,|V|}} = \ket{n_{j,1},\dots,n_{j,i} + d,\dots,n_{j,|V|}} \label{newaji}
    \end{aligned}
\end{equation}

\subsection{Add and delete point}\label{section:Add and delete point}

When there is a need to add or delete point in graph, the dimension of the adjacency matrix will be changed. So as the length of state vector. 
For example, if we simply want to add a new point without any edges adjacent to it in the graph, we need to expand the state vector from $\ket{n_{i,1},n_{i,2},\dots,n_{i,|V|}}$ to $\ket{n_{i,1},n_{i,2},\dots,n_{i,|V|},n_{i,|V|+1}}$, $\forall i \in V$ and $n_{i,|V|+1} =0 $. 
Moreover, we should add a fresh state vector $\ket{n_{|V|+1,1},n_{|V|+1,2},\dots,n_{|V|+1,|V|},n_{|V|+1,|V|+1}}$, definitely, $n_{|V|+1,i} = 0, \quad\forall i=1,2,\dots,|V|+1$. Sure, if we need to add edges adjacent to it, then we can use the operations introduced above in Section \ref{section:Delete and add edge}. 

For deleting point $v_{m}$ in graph, from the aspect of mathematical form, we can simply remove occupation number with the subscript related to $m$ in each row, such as $n_{m,*}$ or $n_{*,m}$. In real quantum systems, however, we need to wash out useless single-particle states. 
The pure process consists of two parts. One is cleaning the $m$th state vector. In Eq. \ref{clean mth}, $ADJ(m)$ is the set of vertices which are adjacent to $m$. After cleaning, all the occupation numbers in $\ket{n_{m,1}^{\prime},n_{m,2}^{\prime},\dots,n_{m,|V|}^{\prime}}$ is 0.
\begin{equation}
  (\prod\limits_{j \in ADJ(m)}a_{m,j}^{n_{m,j}})\ket{n_{m,1},n_{m,2},\dots,n_{m,|V|}} 
= (\prod\limits_{j \in ADJ(m)}(n_{m,j}!)^{1/2}) \ket{n_{m,1}^{\prime},n_{m,2}^{\prime},\dots,n_{m,|V|}^{\prime}}  \label{clean mth}
\end{equation}
Another part is cleaning the remaining $|ADJ(m)|$ vectors by decreasing the occupation number related to $m$ to 0. In Eq. \ref{clean adj}, $DE_{n_{j,m}}^{j,m} = \frac{(a_{j,m})^{d}}{(n_{j,m}!/(n_{j,m}-n_{j,m})!)^{1/2}} = \frac{(a_{j,m})^{d}}{(n_{j,m}!)^{1/2}}$.
\begin{equation}
  DE_{n_{j,m}}^{j,m} \ket{n_{j,1},n_{j,2},\dots,n_{j,m},\dots,n_{i,|V|}} =  \ket{n_{j,1},n_{j,2},\dots,n_{j,m}=0,\dots,n_{i,|V|}}, \quad \forall j \in ADJ(m)  \label{clean adj}
\end{equation}

After these two steps are finished, we can remove the occupation numbers related to $m$ from the $|V| - 1$ vectors.

\subsection{Vertex contraction}

\begin{figure}[H]
  \includegraphics[width=8cm]{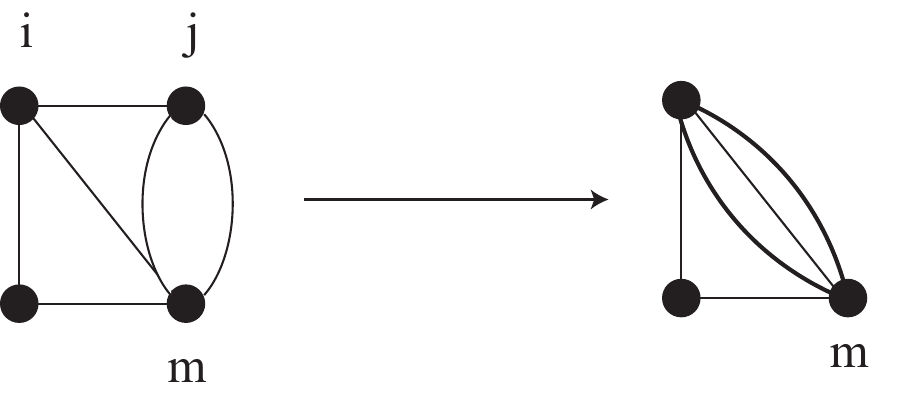}
  \centering
  \caption {An illustration of vertex contraction}
  \label{fig:ss}
\end{figure}

\begin{figure}[H]
  \includegraphics[width=8cm]{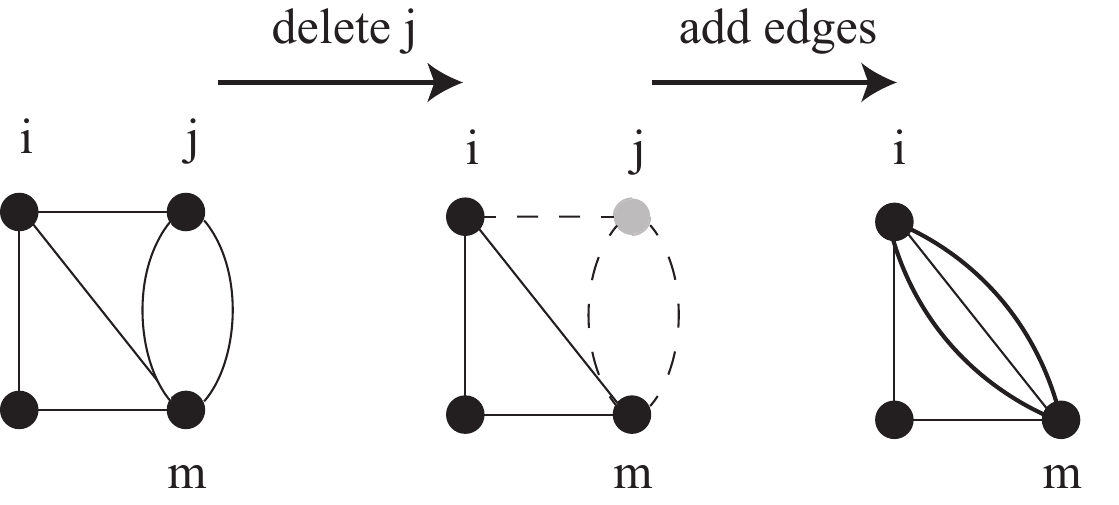}
  \centering
  \caption {Process decomposition of edge contraction}
  \label{fig:ssprocess}
\end{figure}

The operations of adding and deleting edges and vertices introduced in Section \ref{section:Delete and add edge} and \ref{section:Add and delete point} are the fundamental tools in graph transformation. 
Obviously, all complex operations in a graph can be decomposed into operations on edges and vertices. Here, let us consider a somewhat more complicated operation, vertex contraction. 
The contraction of vertices $i$ and $j$ can be described as using a new vertex replacing them, then let the fresh vertex be adjacent to the vertices which are adjacent to $i$ or $j$ such as Fig.~\ref{fig:ss}. 
Vertex contraction has huge applications in graph theory which had been used in many proofs of theorems such as Menger's Theorem \citep{menger1927allgemeinen}, Tutte's Theorem \citep{tutte1947factorization} and so on.
Returning to operations, the contraction process can be decomposed into two processes, deleting one of the vertex and adding the edges adjacent to the deleting one to the left vertex. The process is shown in Fig.~\ref{fig:ssprocess}.

Follow the process in Fig.~\ref{fig:ssprocess}, we can also realize the operation in state vector. First, we can use the operations in Section \ref{section:Delete and add edge} and \ref{section:Add and delete point} to delete $i$ and the edges which are adjacent to $j$. 
This part will not be repeated here. Second, we can let $i$ be adjacent to the vertices which are adjacent to $j$ before. Let $ne_{m,j}, \forall m \in ADJ(j), m\neq i$ be the number of edges between $m$ and $j$. 
Then we can use Eq. \ref{ad im} and \ref{ad mi} to revise the state vectors. This completes the whole operation. 
\begin{equation}
  AE_{ne_{m,j}}^{i,m}\ket{n_{i,1},n_{i,2},\dots,n_{i,m},\dots,n_{i,|V|}} = \ket{n_{i,1},n_{i,2},\dots,n_{i,m} + ne_{m,j},\dots,n_{i,|V|}},\quad  \forall m \in ADJ(j), m\neq i\label{ad im}
\end{equation}
\begin{equation}
  AE_{ne_{m,j}}^{m,i} \ket{n_{m,1},n_{m,2},\dots,n_{m,i},\dots,n_{m,|V|}} = \ket{n_{m,1},n_{m,2},\dots,n_{m,i} + ne_{m,j},\dots,n_{m,|V|}},\quad  \forall m \in ADJ(j),m\neq i\label{ad mi}
\end{equation}

\section{Conclusion}
In this paper, we use the state vector in the occupation number representation to represent the adjacency matrix of a graph. This approach allows us to avoid considering pair-particle interactions. 
Based on the creation and annihilation operators, we can implement fundamental and advanced operations in the graph. 
The generality of our method implies that it can be applied to directed/undirected graphs as well as simple/multigraphs. 
The representation method presented here enriches graph representation theory. 

\section{Author Declaration}
\subsection{Conflict of Interest Statement}
The authors have no conflicts to disclose.
\subsection{Author Contributions}
Haoqian Pan: Writing - original draft; Methodology; Software; Visualization; Investigation. 
Changhong Lu: Writing - review and editing; Methodology; Project administration; Investigation. 
Ben Yang: Investigation.
\subsection{Data Availability Statement}
The data that supports the findings of this study are available within the article.



%
%

%


\bibliography{gr}

\end{document}